\documentclass{ws-ijgmmp}

\usepackage{graphicx}
\usepackage{color}
 \input{epsf}

\def\al{\alpha} \def\be{\beta} \def\ga{\gamma} \def\de{\delta}
\def\ep{\epsilon}   
\def\th{\theta}   \def\ka{\kappa}
   
\def\si{\sigma}   
   
  \def\De{\Delta} 
 \def\Si{\Sigma}  
 \def\Om{\Omega} \def\mn{{\mu\nu}} \def\cl{{\cal L}}

 \def\frac#1#2{{\textstyle{{#1}\over
{#2}}}} 
\def\lsim{\mathrel{\rlap{\lower4pt\hbox{\hskip1pt$\sim$}}
\raise1pt\hbox{$<$}}}
\def\gsim{\mathrel{\rlap{\lower4pt\hbox{\hskip1pt$\sim$}}
\raise1pt\hbox{$>$}}} \def\sqr#1#2{{\vcenter{\vbox{\hrule height.#2pt
\hbox{\vrule width.#2pt height#1pt \kern#1pt \vrule width.#2pt} \hrule
height.#2pt}}}}
\def\square{\mathchoice\sqr66\sqr66\sqr{2.1}3\sqr{1.5}3}
\def\beq{\begin{equation}} \def\eeq{\end{equation}}
\def\beqa{\begin{eqnarray}} \def\eeqa{\end{eqnarray}}
\def\eq#1{Eq. (\ref{#1})}

\begin{document}

\markboth{O. Bertolami and J. P\'aramos}
{Alternative gravity model with non-minimal coupling between matter and curvature}

%
\catchline{}{}{}{}{}
%

\title{Minimal extension of General Relativity: alternative gravity model with non-minimal coupling between matter and curvature}


\author{Orfeu Bertolami}
\author{Jorge P\'aramos}

\address{Departamento de F\'{\i}sica e Astronomia, Faculdade de Ci\^encias, Universidade do Porto,\\Rua do Campo Alegre 687,
4169-007 Porto, Portugal\\
\email{orfeu.bertolami@fc.up.pt} \email{jorge.paramos@fc.up.pt}}

\maketitle

\begin{history}
\received{(\today)}
\end{history}

\begin{abstract}
We examine an extension of General Relativity with an explicit non-minimal coupling between matter and curvature. The purpose of this work is to present an overview of the implications of the latter to various contexts, ranging from astrophysical matter distributions to a cosmological setting. Various results are discussed, including the impact of this non-minimal coupling on the choice of Lagrangian density, on a mechanism to mimic galactic and cluster dark matter, on the possibility of accounting for the accelerated expansion of the Universe, energy density fluctuations and modifications to post-inflationary reheating. The equivalence between a model exhibiting a non-minimal coupling and multi-scalar-theories is also discussed.
\end{abstract}


\section{Introduction}

The origin and nature of dark energy and dark matter are among the most challenging problems of contemporary cosmology. Research on these dark components of the Universe may be divided into two main possibilities: one assumes Einstein's General Relativity (GR) and thus focuses on the search for the dark matter and dark energy contributions; or, alternatively, one assumes that there are important deviations from GR and that dark energy and dark matter can be evaded partially or entirely.

When addressing dark matter \cite{DM}, the first approach relies on the characterization of additional matter species, arising from extensions to the Standard Model of particles, collectively dubbed as weak-interacting massive particles (WIMPS) such as, for instance, neutralinos or axions.

Likewise, the nature of dark energy usually involves the addition of a new scalar field, which slow-rolls down an adequate potential ---   quintessence \cite{Copeland}. A putative unification of dark matter and energy has also been suggested, by resorting to a scalar field \cite{Rosenfeld} or an exotic equation of state (EOS), such as the generalized Chaplygin gas \cite{Chaplygin1,Chaplygin2,Chaplygin3}.

The second approach follows a quite different route, as it assumes that GR is at fault. Thus, several attempts to generalize GR have emerged: these correspond most often to low-energy modifications of Einstein's theory that should be derived from a yet unknown high-energy, fundamental theory of gravitation. In a cosmological context, modifications of the Friedmann equation to include higher order terms in the energy density $\rho$  have been proposed \cite{Maartens,cardassian,scalar}, as well as considerations of the impact of a van der Waals EOS for matter \cite{vdW}.

Changes to the action functional provide a simple pathway for implementing covariant extensions to GR: a straightforward approach replaces the linear scalar curvature $R$ term in the Einstein-Hilbert action with a function $f(R)$ \cite{fR1,fR2} (see \cite{felice} for a review); more complex dependences may also be explored, as considered in Gauss-Bonnet models \cite{GB}.

These classes of $f(R)$ theories enjoy considerable success in several fronts, including the puzzle of the missing ``dark'' matter in galaxies and clusters \cite{DMfR,ClustersfR}, as well as the nature of ``dark'' energy \cite{capoexp}; the early period of rapid expansion of the universe is described by the Starobinsky inflationary model $f(R)=R + \al R^2$ \cite{Staro}; local Solar system impact and the related parameterized post-Newtonian (PPN) metric formalism have also been studied \cite{PPN}.

Given the promising results arising from $f(R)$ theories, one is naturally tempted to further generalize this model. Thus, another interesting possibility is that the coupling between matter and geometry is non-minimal \cite{Lobo}, as written below:
\beq S = \int \left[ \ka f_1(R) + f_2(R) \mathcal{L} \right] \sqrt{-g} d^4 x~~. \label{model}\eeq
\noindent where $\ka = c^4/16\pi G$, $\cl$ is the matter Lagrangian density and $f_i(R) $ ($i=1,2$) are functions of the scalar curvature. A minimal change in GR would correspond to the choice $f_1(R)=R$.

The purpose of this work is to present an overall view of several developments that arise from this framework in a wide range of scenarios \cite{analogyf2,fluid,mimic,clustersBFP,accexp,preheatingf2,perturbations,solarBMP,f2collapse}. Other results not discussed here include the issue of stellar stability \cite{hydro} and the Newtonian limit \cite{Newtonian}, energy conditions \cite{BS}, time machines and wormhole solutions \cite{wormhole} and the mimicking of a cosmological constant at an astrophysical scale \cite{localCC}, which show the versatility and richness of non-minimally coupled theories.

\section{The model}
\label{section:model}

Variation of the action, \eq{model}, with respect to the metric yields the field equations,
\beq \label{field0} 2\left(\ka F_1+  F_2 \cl \right) G_\mn =  2 \De_\mn \left(\ka F_1 + F_2 \cl\right)  - [ \ka( F_1R - f_1 ) + F_2 \cl R] g_\mn + f_2 T_\mn ~~, \eeq
\noindent where $\De_\mn \equiv \nabla_\mu \nabla_\nu - g_\mn \square$ and $F_i(R) \equiv f'_i(R)$. As expected, GR is recovered by setting $f_1(R) = R $ and $f_2(R) = 1$.
By taking the trace of the above, one gets
\beq \label{fieldtrace} \ka (F_1 R - 2f_1) +  F_2 \cl  R =  {1 \over 2} f_2 T - 3\square \left(\ka F_1 + F_2 \cl\right) ~~, \eeq
\noindent where $T$ is the trace of the matter energy-momentum tensor. The usual $f(R)$ theories are recovered by setting $f_2(R) = 1$, so that \eq{fieldtrace} yields an algebraic relation for $R=R(T)$. However, if one considers a non-minimal coupling $f_2(R) \neq 1$, the above becomes a differential equation: in particular, this enables the possibility that $F_2 \cl$ may vary considerably for high curvatures.

\subsection{Non-conservation of energy-momentum}

The Bianchi identities may be used to derive the non-(covariant) conservation of the energy-momentum tensor,
\beq \nabla_\mu T^\mn={F_2 \over f_2}\left(g^\mn \cl-T^\mn\right)\nabla_\mu R~~. \label{cov} \eeq
\noindent In the absence of a non-minimal coupling, $f_2(R)=1$, one recovers the covariant conservation of the energy-momentum tensor. 

\eq{cov} leads to as an extra force imparted on a test particle, 

\beq \label{extraforce}f^{\mu}  =   {1 \over \rho+p}\left[\left(\cl+p\right)\nabla_{\nu}\log f_2 + \nabla_{\nu}p\right]h^{\mu\nu}~~, \eeq

\noindent so that their trajectory will deviate from geodesic motion. Thus, the Equivalence Principle may be violated if the {\it r.h.s.} of the above is non-vanishing.

This is a distinct new feature of theories following \eq{model}: indeed, while in Jordan-Brans-Dicke theories the curvature appears coupled to a scalar field \cite{Damour}, leading to expressions similar to \eq{cov}, this non-conservation may be transformed away by a suitable conformal transformation to the Einstein frame (where the curvature appears uncoupled) \cite{conformal}.

This is not the case of the non-minimally coupled model here discussed: indeed, performing the conformal transformation $ g_\mn \rightarrow \tilde{g}_\mn = f_2 g_\mn $, one finds that the transformed energy-momentum tensor, $\tilde{T}_\mn$, is covariantly conserved only for matter species satisfying $\tilde{T}^\mn = f_2^{-2} T^\mn$ and $ 2\cl = T $ (see Ref. \cite{Sotiriou1} for a thorough discussion) --- in particular, the choices $\cl = -\rho$ and $\cl = p$ \cite{fluid} lead to the EOS $p = -\rho/3$ and $p = \rho$, respectively.

Thus, one concludes that the non-conservation of the energy-momentum tensor \eq{cov} is indeed an intrinsic characteristic of non-minimally coupled theories, instead of an artefact of the choice of frame of reference.

\section{Equivalence with multi-scalar-tensor models}

As mentioned in section \ref{section:model}, through a suitable conformal transformation, the usual $f(R)$ theories can be rewritten as GR with an added scalar field contribution, which is dynamically identified with the curvature, $\phi = R $. Similarly, the discussed non-minimally coupled model \eq{model} can be recast as a multi-scalar field theory, with two scalar fields, albeit a ``physical'' metric remains in the matter Lagrangian \cite{analogyf2}. Indeed, by performing a conformal transformation $g_\mn \rightarrow \tilde{g}_\mn = \exp[(2/\sqrt{3})\varphi^1] g_\mn$, the equivalent action is obtained,
\beq  S =  \int  \sqrt{-\tilde{g}} d^4x \bigg[ f_2(\varphi^2)  e^{-{2 \over \sqrt{3}}\varphi^1} \cl(g_\mn,\chi) + 2 \ka \left( R - 2\tilde{g}^\mn \si_{ij} \varphi^i_{,\mu} \varphi^j_{,\nu} -  4 
U \right) \bigg]~~, \eeq
\noindent where $\chi$ denotes all matter fields, $\varphi^1$ and $\varphi^2$ are scalar fields related to the scalar curvature and matter Lagrangian density through
\beq \varphi^1 = {\sqrt{3}\over2} \log \left[ {F_1(R) + F_2(R) {\cal L} \over 2\ka} \right] ~~, \eeq
\noindent and $\varphi^2 = R $; $\si_{ij}$ is the field-metric
\beq \si_{ij} = \left(\begin{array}{cc}1 & 1 \\ -1 & 0\end{array}\right) ~~,\eeq
\noindent and the potential is given by 
\beq U(\varphi^1,\varphi^2) =  {1 \over 4} \exp \left( -{2 \sqrt{3}\over 3} \varphi^1 \right) \left[\varphi^2 - 
{f_1(\varphi^2 )\over 2\ka }  \exp \left( -{2 \sqrt{3}\over 3} \varphi^1 \right)  \right]~~. \eeq
\noindent Notice that the effective metric in this Lagrangian density (used for contractions in kinetic terms, {\it etc}.) is not $\tilde{g}_\mn$, but instead the metric $g_\mn = \exp[-(2/\sqrt{3})\varphi^1] \tilde{g}_\mn$.

\section{Lagrangian density for a perfect fluid}
\label{section::choice}

The discussion of the preceding section serves to show another striking feature of non-minimally coupled models: the Lagrangian density of matter appears explicitly in the field equations. In particular, one focuses on the case of a perfect fluid, given its ubiquity as a useful description for matter in the Universe. Its energy-momentum tensor has the familiar form
\beq \label{emtensor} T_\mn = (\rho + p)u_\mu u_\nu + p g_\mn ~~,\eeq
\noindent where $\rho$ is the energy density, $p$ the pressure and $u_\mu$ the four-velocity (with $u_\mu u^\mu = -1$); its trace is $T = 3p -\rho$. The associated Lagrangian density is not usually discussed, because it is absent from the field equations of GR: in applications of GR, the choice of Lagrangian density is mostly irrelevant, while it is of the utmost importance when considering a non-minimal coupling \cite{fluid}. 

The identification $\cl = p$ was first advanced in Ref. \cite{Seliger}, with a relativistic generalization in Ref. \cite{Schutz}. Much later, it was shown that this choice is equivalent to $\cl = -\rho$, implemented by a suitable set of thermodynamical potentials and Lagrangian multipliers $\phi^\mu = \varphi_{,\mu}  +s\theta_{,\mu}+\beta_A \alpha^{A}_{,\mu}$; these enable the relativistic thermodynamical relations via a current term $J_\mu \phi^\mu$ in the action, where $J_\mu$ is the vector density, {\it i.e.} the flux vector of the particle number density \cite{Brown}. In Ref. \cite{HE}, an isentropic perfect fluid is described via $\cl = -\rho$.

The equivalence between Lagrangian densities occurs on-shell, by substituting the field equations derived from the matter action
\beq  \label{actionfluid} S_m = \int\left( - \sqrt{-g} \rho + J_\mu \phi^\mu \right) d^4x~~, \eeq
\noindent back into the action functional, leading to the resulting on-shell Lagrangian density $\cl_1 = p$. Similarly, the action may be rewritten so that the on-shell Lagrangian density reads $\cl_2 = na$, where $n = |J|/\sqrt{-g}$ is the particle number density and $a(n,T) = \rho(n)/n - sT$, where $s$ is the entropy per particle and $T$ is the temperature. Lastly, one may remove the current term $J_\mu \phi^\mu/\sqrt{-g} $ from \eq{actionfluid}, thus obtaining the on-shell Lagrangian density $\cl_3 = -\rho$ through the addition of adequate surface terms.

Having this in mind, one now describes how should the procedure could be generalized to a non-minimally coupled scenario. This should affect the terms in \eq{actionfluid} that are minimally coupled, so that the matter action becomes
\beq
  S'_m=\int \left(- \sqrt{-g} f_2(R) \rho + J^\mu \phi_\mu \right) d^4 x~~,
  \label{modified fluid-action}
\eeq
\noindent while the current term remains uncoupled (aside from the use of the metric to contract indexes). Varying the action with respect to the potentials included in $\phi_\mu$, one obtains
\beq - f_2 \mu U_\al = \phi_\al ~~~~,~~~~ T = {1 \over n}{\partial \rho \over \partial s}\Bigg|_n = {1 \over f_2(R)} \th_{,\al}U^\al ~~, \eeq
\noindent where $\mu$ is the chemical potential and $\th$ is a scalar field (included in $\phi_\mu$) whose equation of motion imposes the entropy exchange constraint $(s J^\mu)_{,\mu} = 0$. Thus, the non-minimal coupling of curvature to matter is reflected in both the velocity and the temperature identifications.

By substituting the modified equations of motion into action (\ref{modified fluid-action}), an on-shell Lagrangian density $\cl_1 = p$ may be read, as in GR. The addition of surface integrals also yields the discussed $\cl_2 = -\rho$ and $\cl_3 = -na$.

Even though the action may adopt distinct on-shell forms, this does not imply an equivalence between them: only the original bare Lagrangian density $\cl_0$ should be inserted into the field equations \eq{field0}. However, this bare $\cl_0$ should not appear into the non-conservation law: indeed, when deriving \eq{cov}, the current term (which is not coupled with the metric) is dropped and one is indeed left with $\cl_2 = -\rho$.

Given these considerations, one can adopt a simpler stance regarding the choice of the Lagrangian density $\cl = -\rho$ instead of $\cl = p$: if a dust distribution is to be considered --- that is, a perfect fluid with negligible pressure and corresponding energy-momentum tensor $T_\mn = \rho g_\mn $ ---, it appears unnatural to take a vanishing quantity as Lagrangian density.

Finally, one remarks that if a matter form is actually described by two independent bare Lagrangian densities leading to different dynamical behaviour of Eqs. (\ref{field0}) and (\ref{cov}), then only through direct observation one can be sure about the correct description.

\section{Galactic dark matter mimicking}

One now reviews the mechanism of dark matter mimicking to account  for the flattening of the rotation curves of galaxies \cite{mimic}. In order to isolate the effect of the non-minimal coupling, one sets $f_1(R) = 1$ and assumes a power-law
\beq \label{powerlaw} f_2(R) = 1 + \left({R \over R_n}\right)^n~~. \eeq
\noindent Since the dark matter contribution is dominant at large distances {\it i.e.} low curvatures, a negative exponent $n$ is expected. 

Inserting this into the field \eq{field0}, together with the Lagrangian density $\cl= -\rho$ for a dust distribution with $p=0$, yields
\beqa \label{field0mimic} && \left[ 1 - n \left( {R \over R_n} \right)^n  { \rho \over \ka R} \right] R_\mn - {1 \over 2} R g_\mn =  \\ \nonumber && \left[ 1 + \left( {R \over R_n} \right)^n \right] {\rho \over 2\ka} U_\mu U_\nu - n \De_\mn \left[ \left( {R \over R_n} \right)^n {\rho \over \ka R} \right] ~~. \eeqa
\noindent At large distances, normal matter is subdominant and the trace \eq{fieldtrace} reads
\beq \label{fieldtracemimic2} R  = (1-2n) \left( {R \over R_n} \right)^n  {\rho \over 2\ka }  -3n\square \left[ \left( {R \over R_n} \right)^n  {\rho \over \ka R} \right] ~~. \eeq
\noindent An exact solution is obtained if the last term vanishes:
\beq \label{exact} R = R_n \left[(1-2n) {\rho \over \rho_n}\right]^{1/(1-n)}~~, \eeq
\noindent defining the characteristic density $\rho_n = 2 \ka R_n$.

A more evolved study of \eq{fieldtracemimic2} shows that the most general solution oscillates around the one above: this makes the gradient term actually dominate \eq{fieldtracemimic2}, and also allows for a perturbative solution $f_2(R) \sim 1$. As numerical results show, one may disregard these oscillations and simply consider \eq{exact} \cite{mimic}. A perturbative non-minimal coupling is paramount: it makes the mimicking mechanism satisfy both that the weak, strong, null and dominant energy conditions; grants immunity against Dolgov-Kawasaki instabilities \cite{DK} (see also Ref. \cite{BS}); and turns the extra force arising from \eq{cov} very small.

Now, instead of solving \eq{field0mimic}, obtaining a modified gravitational potential and then reading the mimicked dark matter contribution, one interprets the additional curvature obtained from the non-minimal coupling as due to the density profile of the latter,
\beq \label{mimiceq} \rho_{dm} \equiv 2\ka R = \rho_n \left[(1-2n) {\rho \over \rho_n}\right]^{1/(1-n)} ~~. \eeq
\noindent Thus, one gets a relationship between a visible matter profile $\rho$ and the mimicked dark matter distribution $\rho_{dm}$, with $n$ related to the large radius behaviour of both contributions.

The mimicked dark matter may be further characterized by inserting \eq{exact} into \eq{field0mimic}, and reading the obtained terms as due to a corresponding energy-momentum tensor. The resulting EOS parameter for ``dark matter'',
\beq \label{EOSmimic} w = {p_{dm}  \over \rho_{dm}} = {n \over  1-n} ~~, \eeq
\noindent is then obtained. For a negative exponent $n$, one has $w < 0 $, hinting at at possible cosmological role in enabling an accelerated expansion of the Universe, as will be discussed in a following section.

\begin{figure} 

\epsfxsize=\columnwidth \epsffile{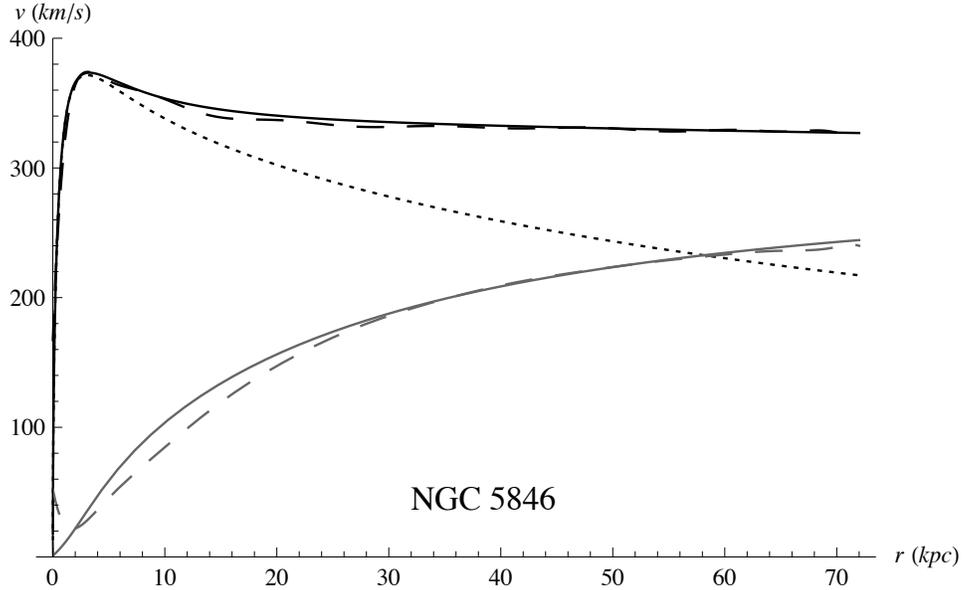}
\caption{Observed rotation curve (dashed full), decomposed into visible (dotted) and dark matter (dashed grey) contributions, with mimicked dark matter profile (full grey) and resulting full rotation curve (full).}
\label{rotation}

\end{figure}

In order to fit the galaxy rotation curves of several galaxies (NGC 2434, 5846, 6703, 7145, 7192, 7507 and 7626, which are almost spherical E0 type galaxies with well determined rotation curves \cite{kronawitter}), one resorts to the Hernquist profile for visible matter \cite{Hernquist}, which behaves as
\beq \label{Hernquistprofile} \rho(r) \sim {a \over r} \left(1+{r \over a} \right)^{-3}~~ , \eeq
\noindent and the Navarro-Frenk-White \cite{NFW} and isothermal sphere profiles for mimicked dark matter: the latter enables a perfectly flat rotation curve, while the former is favored by numerical simulations:
\beq \label{dmprofiles} \rho_{IS}(r) \sim r^{-2}~~~~,~~~~\rho_{NFW}(r) \sim {a \over r} \left(1 + {r \over a} \right)^{-2} ~~, \eeq
\noindent with $a$ signalling the transition between inner and outer slope profiles. Given the outer slope behaviour $\rho(r) \sim r^{-4}$, $\rho_{IS} (r) \sim r^{-2}$ and $\rho_{NFW}(r) \sim r^{-3}$, the relation $\rho_{dm}(r) \propto \rho(r)^{1/(1-n)}$ yields the exponents $n_{IS} = -1$ and $n_{NFW}= -1/3$ that translate the Hernquist profile into the two above. Thus, one is led to consider the composite non-minimal coupling
\beq f_2(R) = 1 + {R_{-1} \over R} + \sqrt[3]{R_{-1/3} \over R}~~. \eeq
The values of the model parameters $R_{-1}$ and $R_{-1/3}$ are obtained from fitting the result of the numerical integration of \eq{field0mimic} to the available galaxy rotation curves. From this, one concludes that $R_{-1} \sim 1/(16.8~Gpc)^2$ and $R_{-1/3} \sim 1/(1.45 \times 10^6 ~Gpc)^2$. Fig. \ref{rotation} shows the fit to NGC 5846 \cite{mimic}.

Allowing for individual fits of these parameters to each of the considered galaxies shows that there is some dispersion, which can be attributed to deviations from sphericity, poor choice of visible or dark matter density profiles, unaccounted effect of a non-trivial $f_1(R) \neq R$ term or the non-minimal coupling with the electromagnetic sector (thus enabling a dependence on luminosity), {\it etc.}. Nevertheless, the quality of the obtained fits and the elegance of the mimicking mechanism provides a robust example of its ability to account for the flattening of the galaxy rotation curves.

\section{Cluster dark matter mimicking}

The mimicking mechanism described in the previous section can also be suitably applied to the description of dark matter at the cluster scale, through the application of \eq{mimiceq}. Ref. \cite{clustersBFP} studies the high-quality data of the Chandra sample of galaxy clusters (A133, A262, A383, A478, A907, A1413, A1795, A1991, A2029, A2390, RX J1159+5531, MKW 4, USGC S152 and A586), which are almost virialized and spherical clusters \cite{vikhlinin}. The density profile for the visible matter present in the cluster assumed to be given by a generalized NFW model,

\beq \rho = \rho_0 {(r/r_c)^{-\alpha}\over (1+r^2/r_c^2)^{3\beta-\alpha/2}}~~, \label{generalized_beta} \eeq

\noindent so that $\rho_g\sim (r/r_c)^{-\alpha}$, in inner regions, and $\rho_g\sim (r/r_c)^{-3\beta}$ for outer regions.

Notice that, contrary to the case for galaxies, in clusters one does not have a dominance of visible matter over dark matter for inner regions; as such, one cannot simply use \eq{mimiceq} to establish a link between the outer slope of their densities, but has also to consider its inner behaviour (for $r \ll r_c$). This indicates that, for each cluster, two exponents $n$ of the power-law non-minimal coupling \eq{powerlaw} are obtained from \eq{mimiceq}, translating either the inner or outer slope of the visible matter density profile into that of dark matter.

Furthermore, the value of the parameter $n$ should be the same for all clusters: thus, one must perform a simultaneous fit of the observed dark matter component of all clusters to the density profile obtained from \eq{mimiceq} for a single $n$.

\begin{figure*}

\epsfxsize=\columnwidth \epsffile{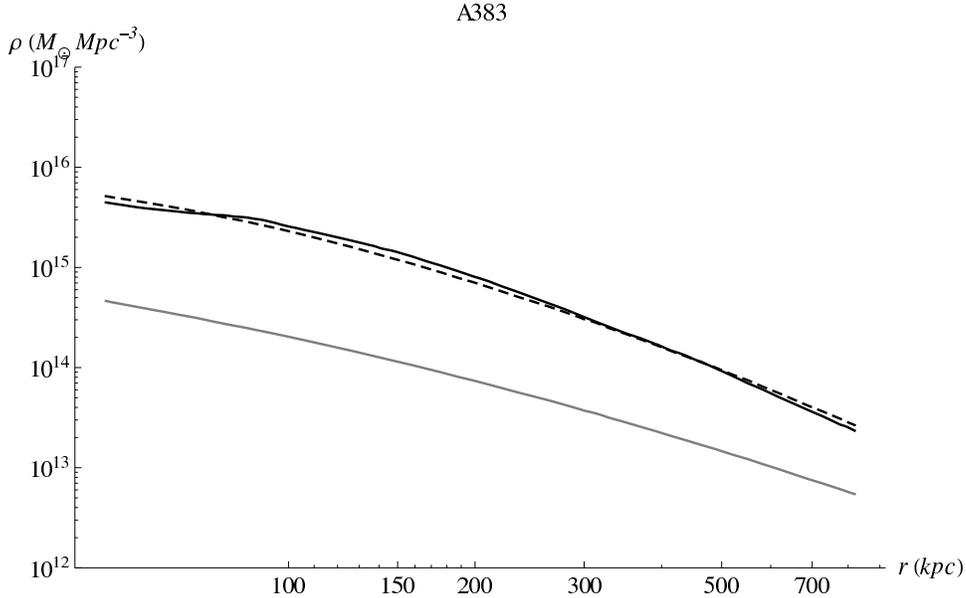}
\caption{The obtained dark matter mimicked profile using a power-law function fixed at $n=0.2$ (dashed), along with the density profiles for dark matter (black) and visible (gray) components for the A383 cluster.}
  \label{mimic1}

\end{figure*}

Numerically, one finds that the best fit occurs for $n= 0.2$, as depicted in Fig. (\ref{mimic1}) for the Abell 383 cluster, here chosen for illustration. Once again, one highlights that the dominance of dark matter at all distances implies that the exponent $n$ does not have to be negative --- contrary to what occurs in galactic dark matter, where $n<0$ was required in order to account for the short-range dominance of visible matter over the latter. Also, notice that \eq{EOSmimic} yields a positive EOS parameter $w= n/(1-n) = 0.25$.

In Ref. \cite{clustersBFP}, it is shown that the mimicked dark matter profile does not depend on the characteristic lengthscale $r_n$, provided that the non-minimal coupling Eq. (\ref{powerlaw}) is perturbative, $f_2(R) \approx 1$ --- leading to the bound $r_{0.2}\lesssim 0.1~{\rm kpc}$. The more stringent bound, $r_n \lesssim 10^{-13}~{\rm kpc}$, is obtained by demanding that the effect of the non-minimal coupling does not lead to an extra force (derived from the non-conservation of the energy-momentum tensor of matte) larger than the Newtonian force. The latter constraint and the best-fit exponent $n=0.2$ yield a perturbative non-minimal coupling $f_2(R) \sim 1 + 10^{-8}$. 

The independence of the mimicked dark matter profile on the lengthscale $r_n$ stems from the fact that \eq{fieldtracemimic2} can be recast in a dimensionless form, such that the latter is effectively absorbed and does not impact the obtained numerical solution. Physically, this reflects the fact that there is no physical lengthscale against which $r_n$ can be gauged, as dark matter dominates at all scales, instead of only above a particular distance (such as $r_c$).

Both the depicted A383 cluster and other clusters studied in Ref. \cite{clustersBFP} present smooth density profiles for visible and dark matter; as such, the mimicking mechanism presented in the previous section can account for the latter in a very satisfactory manner.

\section{Post-inflationary reheating}
\label{section::inflation}

In this section one discusses the possibility of using a non-minimal coupling to drive the reheating of the Universe after inflation \cite{preheatingf2}. One does not attempt to drive the dynamics of inflation itself through the effect of $f_2(R)$, but instead assumes the well-known Starobinsky inflation model \cite{Staro}--- which resorts to a quadratic curvature term
\beq f_1(R) = R + {R^2 \over 6M^2}~~,\label{staroform}\eeq
\noindent with $M \sim 3 \times 10^{-6}M_P$.

However, a non-minimal coupling is fundamental in the so-called preheating mechanism: the reheating of the ultracold post-inflationary Universe due to the explosive production of particles, occurring when the quantum scalar field $\chi$, endowed with a varying mass term, $m^2_{eff} = m^2 + \xi R$, experiences parametric resonance \cite{preheating}. More evolved couplings also lead to preheating, as found in Refs. \cite{mimoso,kinetic}.

This hints that the non-minimally coupled action (\ref{model}) may generalize the preheating scenario. Given the form of the variable mass term $m_{eff}$, one assumes a linear coupling 
\beq f_2(R) = 1 + 2\xi {R \over M^2}~~. \eeq
Since the curvature is  coupled to matter and radiation, besides the scalar field $\chi$, one must ensure that the cosmological dynamics is driven by the effect of the quadratic curvature term \eq{staroform} alone --- {\it i.e.} that the non-minimal coupling intervenes solely during preheating. As shown in Ref. \cite{preheatingf2}, this implies a perturbative regime $f_2(R) \sim 1$ and that $1 < \xi  < 10^4$, compatible with the weak bound $\xi \ll 10^{78}$ obtained from considerations on solar hydrostatic equilibrium \cite{hydro}.

Decomposing the scalar field $\chi$ into its Fourier modes $\chi_k$, one finds that they follow the differential equation
\beqa \label{equationfourier} X''_k + \bigg[\left({ 2 k \over a M}\right)^2 + \left({2m \over M}\right)^2 - 3 {H' \over M} - 9 {H^2 \over M^2} + \\ \nonumber {\xi \over M^2} \left( \xi {R'^2 \over M^2} - 6 {H R' \over M}- R'' \right) \bigg] X_k = 0~~. \label{hilleq} \eeqa
\noindent using the redefinition
\beq X_k \equiv a^{3/2} f_2^{1/2} \chi_k \sim a^{3/2} \chi_k~~,\eeq
where the primes denote differentiation with respect to the variable $z$ (depending on the sign of $\xi$), defined as
\beq 2z = M(t-t_o) \pm {\pi \over 2}~~.\eeq

After slow-roll, the Hubble parameter $H(t)$ experiences an oscillatory phase, so that $R \sim (2M^2 / z) \cos(2z) $; the $R''$ term dominates and one rewrites the above as a Mathieu equation,
\beq X''_k + \left[A_k - 2q \cos(2z) \right] X_k = 0~~, \label{mathieu} \eeq
\noindent with
\beq A_k = \left({2k \over a M}\right)^2 + \left(2 m \over M\right)^2 ~~~~, ~~~~ q = {4 \xi \over z}~~.\eeq
\eq{mathieu} is the same form encountered in usual preheating \cite{preheating}: the quantum field $\chi$ experiences parametric resonance as the scale factor $a(t)$ increases, with massless particles produced for a coupling parameter as low as $\xi \gtrsim 3$, while massive particles require $\xi \gtrsim 10$ --- well within the range $ 1 < \xi < 10^4$ discussed above. Thus, one finds that a universal non-minimal coupling can successfully drive the reheating of the post-inflationary Universe.

\section{Accelerated expansion of the Universe}
\label{section::accelerated} 

This section discusses the possibility of using a non-minimal coupling to describe the current phase of accelerated expansion of the Universe \cite{accexp}. One focuses on a constant deceleration parameter $q \equiv -\ddot{a}a / (\dot{a})^2 = 1/\be - 1$, which translates into a power-law expansion with the scale factor evolving as $a(t) = a_0 (t/t_0)^\be$, with $t_0 = 13.73 ~Gy$.

This suggests the use of a power-law for the non-minimal coupling too (see previous section). Thus, one again considers \eq{powerlaw}, with the exponent $n$ assumed to be negative --- so that the accelerated expansion phase appears at late times, when $R \ll R_n$.

This accelerated expansion is obtained in a rather straightforward fashion: assuming a flat Friedmann-Robertson-Walker (FRW) metric with line element $ ds^2 = - dt^2 + a^2dV^2 $ and that matter is described by a perfect fluid with density $\rho$ and pressure $p$, one has $ T_{00} = \rho $ and $T_{rr} = a^2 p$. One then uses $\cl= -\rho$ and \eq{cov} to ascertain that energy is covariantly conserved in a cosmological context,
\beq \nabla_\mu T^{\mu 0} = {F_2 \over f_2}\left(g^{0 0} \cl-T^{0 0}\right)\dot{R} = 0 \rightarrow\dot{\rho} +3H\rho = 0 \rightarrow \rho(t) = \rho_0 \left({t_0\over t}\right)^{3\be}~~. \label{covcosmo} \eeq
\noindent where $\rho_0 = \Om_m \rho_{crit}$, with $\Om_m \sim 0.3$ the relative matter density and $\rho_{crit}  \sim 10^{-26}~kg/m^3 $ the critical density \cite{WMAP7}.

One uses this result together with \eq{field0} to compute the modified Friedmann equation
\beq \label{modFriedmann} H^2 = {1 \over 6\ka} (\rho + p + \rho_c + p_c)~~, \eeq
\noindent where the additional density $\rho_c$ and pressure $p_c$ terms are introduced,
\beqa \rho_c & =& -6 \rho_0 \be {  1-2\be + n(5 \be + 2n -3)  \over  \left({t\over t_0}\right)^{3\be} \left({t \over t_2}\right)^{2n} \left[6\be (2\be-1)\right]^{1-n} } ~~, \\ \nonumber p_c &=& -2 \rho_0 n { 2+4n^2 -\be (2+3\be) +n(8\be -6) \over  \left({t\over t_0}\right)^{3\be} \left({t \over t_2}\right)^{2n} \left[6\be (2\be-1)\right]^{1-n} }~~.   \eeqa
\noindent after defining $t_n \equiv R_n^{-1/2}$ in the weak regime $F_2 \rho \ll \ka$ \cite{accexp}.

\begin{figure} 

\epsfxsize=\columnwidth \epsffile{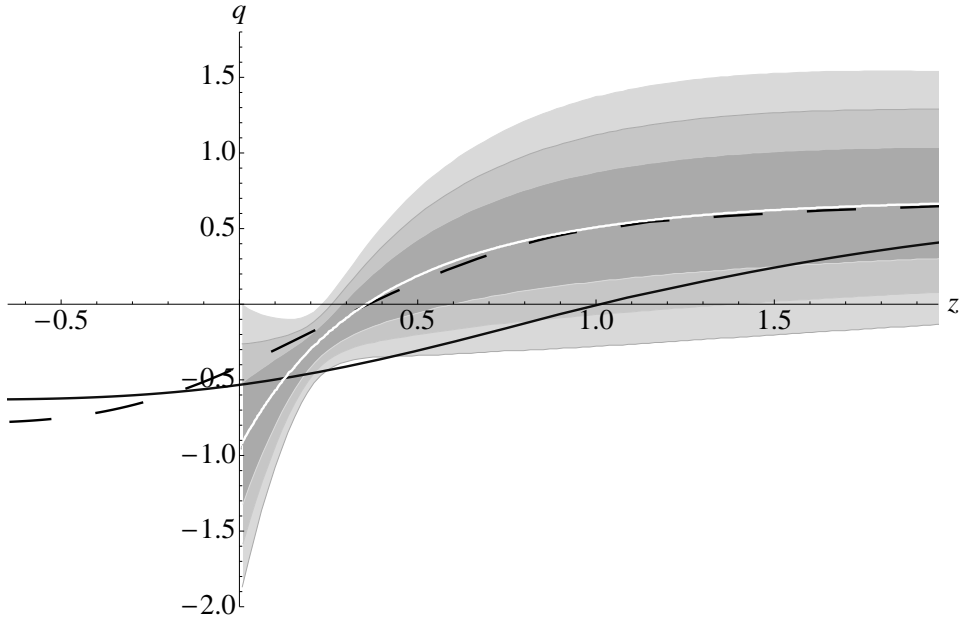}
\caption{Evolution of the deceleration parameter $q(z)$ for $n=-4$, $t_2=t_0/4$ (full) and $n=-10$, $t_2=t_0/2$ (dashed); $1\si$, $2\si$ and $3\si$ allowed regions are shaded, white line gives best fit ({\it cf}. Ref. [X]).}
\label{qzgraph}

\end{figure}

Given the Hubble parameter 
\beq H(t) \equiv {\dot{a} \over a} = {\be\over t} ~~, \eeq
\noindent the {\it l.h.s.} of the Friedmann \eq{modFriedmann} falls as $t^{-2}$, so that comparing with the above gives a relation between the exponents $\be$ and $n$:
\beq 3\be + 2n = 2 \rightarrow \be = {2 \over 3}(1-n)~~. \eeq
\noindent Thus, one concludes that any negative exponent $n$ will yield an accelerating Universe for which $\be > 0$ and $q < 0$ at late times. One may also compute the deceleration parameter and EOS parameter for the non-minimally coupled contribution,
\beq q = - 1 + {3 \over 2(1-n)} ~~~~,~~~~w = {n \over 1 - n}~~. \label{eqqz} \eeq
\noindent The latter is the same form found in the previous section. By fitting the numerical solution of \eq{modFriedmann} to the evolution profile of $q(z)$ \cite{gong}, one obtains the best fit values $n = -10$ and $t_{-10} = t_0/2 $, depicted in Fig. \ref{qzgraph}.

This value for the exponent $n$ is very distinct from the $n_{IS} = -1$ and $n_{NFW} = -1/3$ scenarios considered in the previous section. However, these have no cosmological impact on the current scenario, given the broad difference between the relevant timescale $t_0$ and those obtained from $t_{-1} \equiv R_{-1}^{-1/2}$ and $t_{-1/3} \equiv R_{-1/3}^{-1/2}$. The value $n = -10$ may also be regarded as somewhat unnatural, but is highly dependent on the $q(z)$ profile: Fig. \ref{qzgraph} shows that $n = -4$, $t_2 = t_0/4$ also yields a curve within the allowed $3\si$ region. Thus, despite this caveat, one concludes that a non-minimal coupling may be used to describe the accelerated expansion of the Universe.

\section{Cosmological Perturbations}

In this section one ascertains how the presence of non-trivial functions $f_1(R)$ and $f_2(R)$ affects the evolution of cosmological perturbations --- so that any deviation from the usual GR dynamics can be used to probe the latter \cite{perturbations}. Indeed, it is known that $f(R)$ theories can leave characteristic imprints on the matter power spectrum \cite{pert11,pert12,pert13,pert14,pert15} and modify gravitational lensing \cite{Zhang}.

Cosmological perturbations can be decomposed into scalar, vector and tensorial modes \cite{Mukhanov}; however, since vector modes decay with the cosmological expansion and the tensorial ones are decoupled from the energy density or pressure because the non-diagonal terms of the energy-momentum tensor of matter are zero, one focuses solely on scalar perturbations --- which grow during the expansion of the Universe and lead to the formation of large scale structure (LSS). Thus, in the longitudinal gauge \cite{Mukhanov}, the perturbed FRW metric is

\begin{equation}
ds^{2} = -(1+2\Phi)\, dt^{2}+a^{2}(t)(1-2\Psi)\tilde{g}_{ij}dx^{i}\, dx^{j}~~.
\end{equation}

One obtains the equations ruling the evolution of cosmological perturbations by linearizing the modified field \eq{field0} and non-conservation of the energy-momentum tensor \eq{cov}. In particular, the non-diagonal components of the former yield the relation

\beq \Phi - \Psi = - \de \log \left( F_1 - {F_2 \rho\over \ka} \right)~~, \eeq

\noindent which should be contrasted with the identity $\Phi = \Psi$ obtained in GR.

The relevant scale for the formation of LSS corresponds to modes deep inside the horizon during the matter dominated era, for which $k^{2}/a^{2}H^{2}\gg 1$. Neglecting time derivatives of the potentials and keeping only terms on the energy density fluctuations \cite{Cruz}, one can show from the linearized field \eq{field0} that the generalized Newtonian potential $\Phi$ obeys a modified Poisson equation,

\begin{equation}
\Phi\approx-4\pi \tilde{G}{a^2\over k^2}\delta \rho ~~,
\end{equation}

\noindent with the modified gravitational constant $\tilde{G}$ given by 

\begin{equation}
 {\tilde{G}\over G}\equiv {1+2 {k^{2}\over a^{2}R}\left(2m_{1}-m_{2}\right)\over 1+3 {k^{2}\over a^{2}R}m_{1}}\Sigma~~,
\end{equation}

\noindent introducing the weak lensing parameter $\Sigma$ \cite{weaklensing1,weaklensing2} and quantities $m_1$ and $m_2$,

\begin{equation}\label{defSigma}
\Sigma \equiv { f_2\over F_1- {F_2\rho \over \ka} }~~~~,~~~~ m_{1}\equiv =R{\ka F^\prime_{1}-F^\prime_{2}\rho \over \ka F_{1}-f_2\rho}~~~~,~~~~m_{2} \equiv R{f_2\over f_2}~~.
\end{equation}

\noindent In GR, one has $\Si = 1$, $m_1 =m_2=0$, so that $\tilde{G} = G$ and the usual Poisson equation is recovered.

By the same token, the equation for the second potential has also a Poisson form,

\beq \Psi\approx -4\pi G s{a^{2} \over k^{2}}\delta\rho ~~~~,~~~~ s={1+2{k^{2}\over a^{2}R}\left(m_{1}+m_{2}\right)\over 1+3{k^{2}\over a^{2}R}m_{1}}\Sigma ~~.
\end{equation}

\noindent In GR, one obtains $s=1$, confirming that indeed $\Psi = \Phi$, as expected.

Proceeding along these lines, Ref. \cite{perturbations} shows that the gauge-invariant matter overdensity  $\delta_m\equiv \delta \rho/\rho +3H v$ (where $v$ is the potential velocity, defined through $u_i = - v_{,i}$) obeys the differential equation

\begin{equation}
\ddot{\delta}_{m}+(2H+\dot{\Phi}_c)\dot{\delta}_{m}-4\pi (\tilde{G}+G_c)\rho_m\delta_{m}\simeq 0~~,
\label{delta_m}
\end{equation}

\noindent with the coupling potential $\Phi_c \equiv \log f_2$ acting as a friction force added to the effect of the Hubble expansion, and the coupling

\begin{equation}
{G_c\over G}\equiv {-{k^{2}\over a^{2}R}m_{2}\left(1+6{k^{2}\over a^{2}R}m_{2}\right)\over 1+3 {k^{2}\over a^{2}R}m_{1}}\Sigma~~.
\end{equation}

\noindent Thus, the non-minimal coupling modifies the frictional term and the attractive interaction; depending on the form of $f_2(R)$, these two effects can either reinforce or oppose each other.

One now ascertains how the results presented above behave if the non-minimal coupling has a power-law form \eq{powerlaw} with a negative exponent $n$ --- required to drive the accelerated expansion of the Universe, as discussed in Section \ref{section::accelerated}. One gets

\begin{equation}\label{casesSigma}
\Sigma ={\left( {R \over R_n}\right)^n \over 1-2n{\kappa \rho \over R} \left( {R \over R_n}\right)^n }\approx
\begin{cases}
1+\left(1+2n{\kappa \rho \over R} \right)\left( {R \over R_n}\right)^n  ~~&,~~ \left( {R \over R_n}\right)^n \ll 1 \cr 
  -{1\over 2n}{{R\over \kappa\rho} } ~~&,~~\left( {R \over R_n}\right)^n \gg 1
\end{cases}~~,
\end{equation}

\noindent along with

\begin{eqnarray}\label{regimes}
m_{1}&=&{2n\left(1-n\right)\frac{\kappa\rho_m}{R}\; \left( {R \over R_n}\right)^n \over 1-2n\frac{\kappa\rho_m}{R}\left( {R \over R_n}\right)^n }\approx
\begin{cases}
{2n\left(1-n\right)\frac{\kappa\rho_m}{R}\; \left( {R \over R_n}\right)^n } ~~&,~~\left( {R \over R_n}\right)^n \ll 1 \cr
{n-1} ~~&,~~\left( {R \over R_n}\right)^n \gg 1
\end{cases} \nonumber ~~,\\
m_{2}&=&n{\left( {R \over R_n}\right)^n \over 1+\left( {R \over R_n}\right)^n }\approx
\begin{cases}
n{f_2}  ~~&,~~\left( {R \over R_n}\right)^n  \ll 1 \cr 
 n  ~~&,~~ \left( {R \over R_n}\right)^n \gg 1
\end{cases}~~,
\end{eqnarray}

\noindent and

\begin{eqnarray}
\dot{\Phi}_{c} & =& n{\left( {R \over R_n}\right)^n \over 1+\left( {R \over R_n}\right)^n }{\dot{R}\over R}\approx
\begin{cases}
n\left( {R \over R_n}\right)^n \frac{\dot{R}}{R}~~&,~~ \left( {R \over R_n}\right)^n  \ll 1 \cr
 n\frac{\dot{R}}{R}  ~~&, ~~ \left( {R \over R_n}\right)^n \gg1
\end{cases}~~.
\end{eqnarray}

A negative exponent $n$ implies that the non-minimal coupling is negligible at early times, when the scalar curvature is high, $R\gg R_n$, while becoming more important as the Universe expands and the latter drops. Thus, one sees from \eq{regimes} that the quantities $m_1$ and $m_2$ rise from small values to those of order unity (if $n \sim {\cal O}(1)$) when the accelerated expansion phase of the Universe is underway. One also sees that, since the curvature decreases with time, for $n<0$ the frictional potential $\dot{\Phi}_c$ is positive, thus reinforcing the frictional term due to the Hubble expansion.

In order to see how the matter fluctuations grow, one first ensures that, early in the matter dominated era, the quantities $m_i$ are small enough to enable the condition $k^2m_i \ll a^2 R $: using \eq{regimes}, one has

\begin{equation}
{\tilde{G}+G_c\over G}\approx\left[1-2n\left(1+n\right){k^{2}\over a^{2}R}\left( {R \over R_n}\right)^n \right]\Sigma~~,
\end{equation}

\noindent so that the density perturbation $\de \equiv \de \rho /\rho$ obeys the differential equation,

\begin{equation}\label{difeqdelta}
{d^2\delta \over d^2N}+{1 \over 2}{d\delta \over dN}-{3\over 2}\left[1+A_n(k,a)\right]\delta_{m}=0~~,
\end{equation}

\noindent where $N = \log (a/a_0) $ is the e-fold number and

\begin{equation}
A_n(k,a)\equiv {k^{2}\over a^{2}R}\left(m_{1}-4m_{2}\right)= -2n\left(1+n\right){k^{2}\over a^{2}R} \left( {R \over R_n}\right)^n  \sim a^{1-3n}~~,
\end{equation}
 
\noindent using $a \sim t^{2/3} \rightarrow R \sim H^2 \sim 1/t^2 \sim a^{-3}$ during the matter dominated era.

An approximate solution to \eq{difeqdelta} is given by

\beq \label{growthmatter}\delta \sim a^{1+{6n\left(1+n\right)\over 5\left(3n-1\right)}{k^{2}\over a^{2}R N} \left( {R \over R_n}\right)^n} ~~. \eeq

\noindent In GR, the effect of the non-minimal coupling is absent and one recovers the linear growth $\de \sim a$. Notice that $-1<n<0$ leads to increased growth, and conversely $n< -1$ decreases the generation of density perturbations.

As the curvature drops and the non-minimal coupling becomes more relevant and starts driving the Universe towards a phase of accelerated expansion, the parameters $m_i$ become of order unity and one gets $k^2m_i \gg a^2 R $, so that repeating the steps outlined above yields

\begin{equation}
{\tilde{G} + G_c \over G}\approx{G_{c}\over G}\approx-4{k^{2}\over a^{2}R}{m_{2}^{2}\over m_{1}}\Sigma\approx{2n\over n-1}{k^{2}\over a^{2}}{1\over \kappa}~~,
\end{equation}

\noindent and, since $\rho \sim a^{-3}$, one finds that the interaction increases linearly $\tilde{G} + G_c\sim a$.

At this stage, the dynamics of density perturbations is ruled by

\begin{eqnarray}
{d^2\delta \over dN^2}+{1\over 2}{d\delta \over dN}+B_n(k,a)\,\delta_{m}=0~~,
\end{eqnarray}

\noindent with

\begin{equation}
B_n(k,a)\equiv-6{k^{2}\over a^{2}R} {m_{2}^{2}\over m_{1}}\left( {R \over R_n}\right)^n={6n^{2}\over 1-n} {k^{2}\over a^{2}R}\left( {R \over R_n}\right)^n\sim a^{1-3n}~~.
\end{equation}

An approximate solution to the equation above is 

\begin{equation}\label{growthDE}
\delta\sim a^{{2\over \left(1-3n\right)N}\sqrt{{6n^{2}\over 1-n}{k^{2}\over a^{2}R}\left( {R \over R_n}\right)^n}}~~,
\end{equation}

\noindent yielding an accelerated growth of density perturbations. 

As a conclusion, one highlights that the above results, Eqs. (\ref{growthmatter}) and (\ref{growthDE}), together with the modified weak lensing parameter $\Si$ expressed in Eqs. (\ref{defSigma}), (\ref{casesSigma}), allow in principle for a better discrimination of a putative non-minimal coupling acting as dark energy. Furthermore, a negative exponent $n$ is found to be not only required to drive the late time accelerated expansion of the Universe, but is also required by the constraint $\Sigma >0$. Thus, one concludes that the dark energy model presented in Section \ref{section::accelerated} does not, in principle, lead to an incompatibility with large scale structure formation.

\section{Parameterized Post-Newtonian Formalism}

This section briefly summarizes how one can assess the impact of the non-minimally coupled theory \eq{model} in the Solar System, by following the well-known Parameterized Post-Newtonian (PPN) Formalism. This extremely useful tool is based upon an expansion of the metric elements and other quantities (energy-momentum tensor, equations of motion, {\it etc}.) in powers of $1/c^2$, and successfully translates relevant features of modified gravity theories into a set of so-called PPN parameters ({\it e.g.} violation of momentum conservation, existence of a privileged reference frame, amongst others deviations from GR). For simplicity, one focuses only on the $\be$ and $\ga$ PPN parameters, which measure the amount of non-linearity affecting the superposition law for gravity and the spatial curvature per unit mass, respectively, and yield the PPN metric \cite{Will}

\beq ds^2 = - \left[ 1 + {2V \over c^2} + 2 \be \left({ V \over c^2}\right)^2 \right]~(c~dt)^2 + \left( 1 + 2 \ga {V \over c^2} \right)~\left( dr^2 + d\Om^2 \right) ~~.\eeq

\noindent where $V$ is the gravitational potential.

By definition, General Relativity is parameterized in the PPN formalism by $\be=\ga=1$, while all remaining parameters vanish; measurements of the Nordtvedt effect yield $|\be -1 | \leq 2-3 \times 10^{-4}$ \cite{beta}, while Cassini radiometry indicates that  $\ga -1 = (2.1 \pm 2.3) \times 10^{-5}$ \cite{gamma} (see Refs. \cite{survive,status} for a discussion).

Aiming at describing $f(R)$ theories through the PPN formalism, one assumes that the curvature scalar $R$ can be decomposed into a spatially varying component $R_1(r)$ and a time dependent cosmological background component $R_0(t)$, $R(r,t) = R_0(t) + R_1(r)$ \cite{PPN} (see also Refs. \cite{PPNfR}); $R_0(t)$ is assumed to vary slowly when compared with typical timescale of Solar System phenomena. This leads to the modification of the metric, which is now written as a perturbation to the background Friemann-Roberton-Walker metric,

\beq \label{metric}ds^2 = -\left[1 + 2\Psi(r,t) \right] dt^2 + a^2(t)\left(\left[1 + 2\Phi(r,t)\right] dr^2 + r^2 d\Omega^2 \right)~~, \eeq

\noindent with the scale factor $a(t)$ normalized as $a(t_0)=1$, where $t_0$ is the present time.

A strong constraint derived from the application of the method outlined in Ref. \cite{PPN} is that the modification to the curvature induced by the matter source (the Sun, in the case of the Solar System) is assumed to be perturbative, $R_1(r) \ll R_0(t)$.

In Ref. \cite{solarBMP}, this method was expanded to include the presence of a non-minimal coupling in \eq{model}. Its validity rests upon the set of conditions listed below:

\begin{itemize}
\item Long-range: The additional interaction resulting from non-trivial $f_1(R)$ and $f_2(R)$ functions must be long-ranged, {\it i.e.} the related mass scale

\beqa \label{mass-formula} && m^2 = {1 \over 3}\bigg[  - R_0 + \\ \nonumber && {F_1(R_0) + F_2(R_0)(\rho + \rho_{cos}) - 3\square\left[ F'_1(R_0) - 2 F'_2(R_0)\rho^{\rm cos}\right] + 6\rho \square F'_2 (R_0) \over F'_1(R_0) - 2 F'_2(R_0) (\rho + \rho_{cos})}\bigg]~~, \eeqa

\noindent must be small at Solar System scales, $|mr| \ll 1$, both inside and outside the central body. Here, $\rho(r)$ is the density of the latter and $\rho_{cos}(t)$ is the cosmological background density.

\item Perturbative regime: the condition $R_1(r) \ll R_0(t)$ must be obeyed, so that the first-order expansions of \eq{field0} in a spherical spacetime used to ascertain the PPN parameters $\be$ and $\ga$ are valid. 

\item Newtonian regime: the requirement that the potential $\Psi$ is proportional to $M/r$ (with $M$ the mass of the central body) leads to the condition inside the spherical body

\beq \label{Newt-limit} \left\vert 2 F_2 (R_0) \right\vert \rho(r) \ll \left\vert F_1 (R_0) - 2 F_2 (R_0)\rho_{cos}(t) \right\vert~~.\eeq

\end{itemize}

If all of the above conditions are upheld, then the PPN parameter $\ga$ may be read from

\beq \label{solargamma}\ga  = {1\over 2} { f_2(R_0) + 4 F_2(R_0) R_0 + 12\square F_2(R_0) \over  f_2(R_0) + F_2(R_0) R_0 +3\square F_2(R_0)} ~~.\eeq

\noindent Notice that inserting $f_1(R) = R $ and $f_2(R) = 1$ into the above does not yield the value $\ga=1$ found in GR, as the procedure used to obtain this expression is no longer valid (as can be seen from the vanishing denominator in \eq{mass-formula}). However, for non-trivial forms for these functions, the stringent constraint $\ga -1 = (2.1 \pm 2.3) \times 10^{-5}$ \cite{gamma} should lead to strong constraints on the parameters characterizing $f_1(R)$ and $f_2(R)$, or outright dismiss these \cite{PPN}.

In the present context, t is natural to resort to the model previously described in Section \ref{section::accelerated} as driving the accelerated expansion of the Universe, {\it i.e.} considering a power-law non-minimal coupling \eq{powerlaw} with a negative exponent $n<0$.

Using the expressions obtained in Ref. \cite{accexp} for the cosmological component of the scalar curvature $R_0(t)$ and density $\rho_{cos}(t)$, and taking as density profile for the Sun the expression below \cite{NASAdensity}

\beq \label{NASAprofile} \rho = \rho_{c0} \left[ 1 - 5.74 {r \over R}+ 11.9 \left({r \over R}\right)^2 - 10.5\left({r \over R}\right)^3 + 3.34\left( {r \over R}\right)^4\right] ~~,
 \end{equation}

\noindent where $R$ is the radius of the central body and $\rho_{c0}$ its central density, one may ascertain the validity of the conditions listed before:

\begin{itemize}
\item Long-range regime $|m| r \ll 1$ within the Solar System: leads to $n \gg 10^{-25}$;
\item Newtonian approximation: requires that $n \ll 10^{-33}$;
\item Perturbative regime $\left\vert R_1 \right\vert \ll \left\vert R_0 \right\vert$: one finds that $R_1/R_0 \sim 1/(1+n)$ (unless the density of the central body is fine-tuned in an unphysical fashion so that $R_1/R_0 \ll 1$)) --- which disallows a first-order expansion of relevant quantities around $R= R_0(t)$, as second-order terms turn out to be of the same order of magnitude (due to a cancelling factor $1+n$).
\end{itemize}

Given the incompatibility between the two ranges obtained for the exponent $n$ (and with the best fit values $n=-4,~-10$ \cite{accexp}), and the breaking of the perturbative regime, one concludes that the model posited in Ref. \cite{accexp} cannot be constrained by Solar System observables using the method here summarized. Nevertheless, it provides a useful tool to guide future research into the impact of a cosmologically relevant non-minimal coupling in the Solar System.

\section{Gravitational Collapse}

One now discusses how the gravitational collapse of a spherical body of homogeneous dust is affected by a non-minimal coupling, thus generalizing the familiar Oppenheimer-Snyder scenario (OS) \cite{f2collapse}. In order to obtain a tractable problem, one considers a linear $f_2(R) = 1 + \ep R/\ka$ and a trivial $f_1(R)= R$, highlighting the effect of the former. Compatibility between a non-minimally coupled preheating mechanism and Starobinsky inflation \cite{preheatingf2} dictates that the coupling strength is of the order $10^9 < \ep < 10^{13}$, as discussed in Section \ref{section::inflation}.

Although in Section \ref{section::choice} it was argued that $\cl = - \rho$, not $\cl = p$, is the most adequate description of a perfect fluid \cite{fluid}, one assesses the physical impact of both this form and $\cl = p$ by writing $\cl = - \al \rho$: the former choice corresponds to $\al = 1$, while the latter is obtained by setting $\al = 0$ (as the pressure vanishes in the dust distribution). The parameter $\al$ thus acts as a binary variable, considerably simplifying several calculations.

This, together with the energy-momentum tensor for dust and the adopted forms for $f_1(R)$ and $f_2(R)$, leads to the modified field \eq{field0},

\beq {1 \over 6}\ka a^2 \rho = \left( \ka^2 - \ep \rho \right) k + \left[ \ka^2 +(3\al-1) \ep \rho \right] (\dot{a})^2 + \ep (\al-1) a \ddot{a} \rho~~,  \label{fieldset} \eeq

\noindent where $a(t)$ is the scale factor and $k$ the spatial curvature of the FRW metric describing the homogeneous collapse: the latter deviates from its value $k_0$ in GR through 

\beq k = {k_0 \over 1 - {1+\al \over 2}\ep_0} ~~,\eeq 

\noindent with $\ep_0 \equiv \ep \rho_0/\ka^2$. Assuming that the density of the spherical body before collapse is much smaller than the typical density of neutron stars, $\rho_0 \ll \rho_N \sim 10^{18}~{\rm kg/m}^3$, this  is vanishingly small, $\ep_0 \ll 10^{-62}$.

Using \eq{fieldtrace} and recalling that $\al$ is a binary variable, one may integrate \eq{cov} to find 

\beq \label{noncons} \rho(t) = \rho_0\left({\ka + \ep R \over \ka + \ep R_0}\right)^{ \al - 1 } \left({a_0 \over a}\right)^3 = { \rho_0 a^{-3} \over 1 + \ep (1-\al) {\rho_0 \over 2\ka^2} \left( a^{-3} - 1 \right) }~~. \eeq

Combining \eq{fieldset} and (\ref{noncons}) yields the differential equation

\beq \label{dota} \dot{a} = - \sqrt{k(1-a) { a^2 +  (a + 1) \al \ep_0 \over a^3 + 2 \al \ep_0}}~~~~,~~~~a(t_0)= 1~~.\eeq

\begin{figure} 

\epsfxsize=\columnwidth
\epsffile{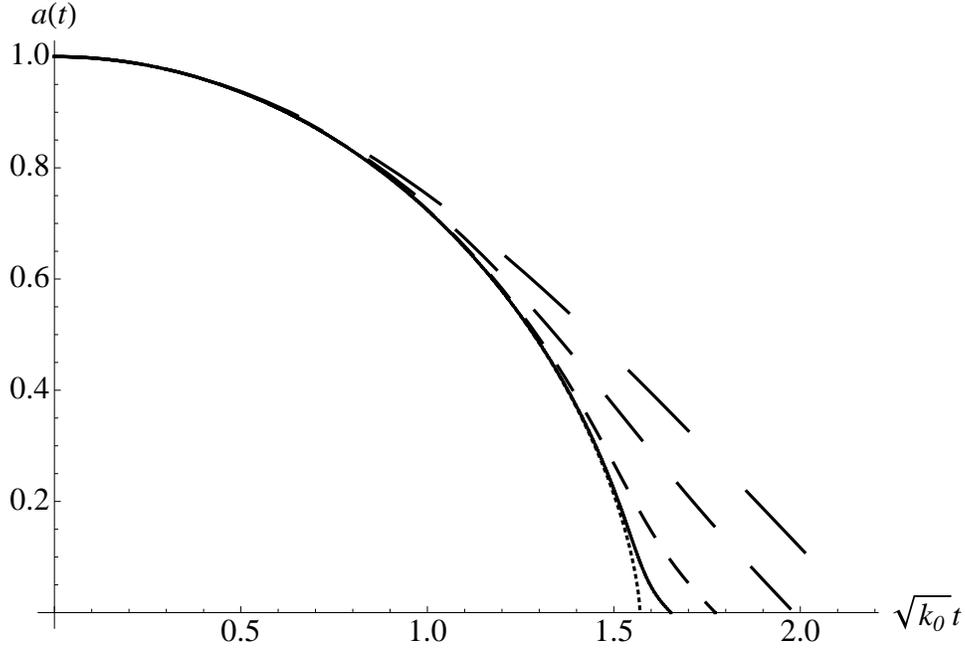}
\caption{Evolution of the scale factor for different values of $\ep_0 = [10^{-3},~10^{-2},~,10^{-1},~1]$ (full line, small, medium and large dash); dotted indicates $\ep_0 =0$.}
\label{figcollapse}

\end{figure}

Eqs. (\ref{dota}) and (\ref{noncons}) shows that, if $\cl = - \rho \rightarrow \al = 1$, the gravitational collapse deviates very weakly from GR; this is shown in Fig. \ref{figcollapse}, with unphysically large values of $\ep_0$ depicted, for clarity. The usual dependence of the density on the scale factor $\rho \sim a^{-3} $ is maintained, and a point-like singularity with infinite density reached.

The $\cl = p \rightarrow \al = 0$ case is much more interesting: although the scale factor evolves qualitatively as in OS collapse (since \eq{dota} only includes the redefinition of the spatial curvature $k$ with respect to its value $k_0$ in GR), the modified dependence for the density, \eq{noncons}, yields a geometric point-like singularity with a final finite density $\rho \rightarrow \rho_f = 2\ka^2/\ep$. Since $\ep$ is very large, this falls well below the Planckian domain, $\rho \rightarrow \rho_f \ll M_P^4$ --- although still many orders of magnitude above the typical density of neutron stars.

To match the interior FRW metric with the outer Schwarzschild metric, one inspects the behaviour of the extrinsic curvature $K_{ab}$ across the surface of the body,

\beq \label{junction2} K^+_{ab} =  \left(1 - {\ep \al \over \ka^2} \rho \right) K^-_{ab} + {\ep \al \over \ka^2} \rho K^- h_{ab} ~~. \eeq

\noindent where $h_{ab}$ is the induced metric on the surface of the collapsing body. 

In the $\cl = p \rightarrow \al = 0$ scenario, the extrinsic curvature is continuous, and the mass $M$ of the spherical body (derived from the gravitational potential in its exterior) deviates from  the usual gravitational mass $M_0$ found in GR through 

\beq M = {M_0 \over 1 - {1 \over 2}\ep_0} ~~.\eeq

\noindent Given the dependence on the initial density $\rho_0 = (\ep/\ep_0) \ka^2$, event horizons at different distances arise from the collapse of stars with the same mass, but distinct radius --- thus breaking the no-hair theorem.

The alternative description $\cl = - \rho \rightarrow \al = 1$ is problematic, since the matching of the inner and outer spacetimes is unfeasible unless unnatural extra terms are added to the boundary action, as thoroughly discussed in Ref. \cite{f2collapse}. This can be related to the non-vanishing effective pressure $p_{eff} \sim G_{rr} \neq 0$ due to the non-minimal coupling, reminiscent of the similar matching problem found in the gravitational collapse of a homogeneous sphere with pressure in GR (see a related discussion in Ref. \cite{Sotiriou3}).

\section{Conclusions and Outlook}

A non-minimal coupling between geometry and matter has a broad range of implications. These suggest that a non-minimal coupling should be a feature of a fundamental quantum theory of gravity.

The analytic expressions of the non-minimal coupling considered may be regarded as an approximation to a more elaborated form for $f_2(R)$, each valid in a particular regime: early {\it vs.} late time, central {\it vs.} long range, {\it etc}.. The latter could perhaps be written as a sum of power-law terms, so that one could chart yet unprobed terms of the above series by examining the dynamics of other phenomena and environments.

Of  course, outstanding challenges lie ahead, such as for instance, the dynamics of active clusters such as the ``bullet cluster'' \cite{bullet}; however, the obtained results are encouraging and, of course, the search for the underlying fundamental quantum gravity theory continues --- and we believe that the requirement for an effective theory that exhibits a non-minimal coupling becomes stronger with the results obtained so far. 

\section*{Acknowledgements}
The work of the authorsf is partially supported by FCT (Funda\c{c}\~ao para a Ci\^encia e a Tecnologia, Portugal) under the project PTDC/FIS/111362/2009. This contribution was developed in the context of the {\it 49$^{th}$ Winter School of Theoretical Physics --- Cosmology and non-equilibrium statistical mechanics}, L\c{a}dek-Zdr\'oj, Poland, February 10-16, 2013. OB wishes to thank the organization for the kind invitation and hospitality. The authors thank P. Fraz\~ao and C. Bastos for fruitful discussions.


\begin{thebibliography}{0}

\bibitem{DM}G. Bertone, D. Hooper, and J. Silk, {\it Physics Reports} {\bf 405} (2005) 279.

\bibitem{Copeland}E. J. Copeland, M. Sami and S. Tsujikawa, {\it International Journal of Modern Physics D} {\bf 15} (2006) 1753.

\bibitem{Rosenfeld}O. Bertolami and R. Rosenfeld, {\it International Journal of Modern Physics} {\bf  A 23} (2008) 4817.

\bibitem{Chaplygin1}A. Kamenshchik, U. Moschella and V. Pasquier, {\it Physics Letters B} {\bf 511} (2001) 265.

\bibitem{Chaplygin2}N. Bili\'c, G. Tupper and R. Viollier, {\it Physics Letters B} {\bf 535} (2002) 17.

\bibitem{Chaplygin3}M. C. Bento, O. Bertolami and A. A. Sen, {\it Physical Review D} {\bf 66} (2002) 043507.

\bibitem{Maartens}R. Maartens, {\it Living Reviews in Relativity} {\bf 7} (2004) 7. 

\bibitem{cardassian}K. Freese and M. Lewis, {\it Physics Letters B} {\bf 540} (2002) 1.
  
\bibitem{scalar}S. Carroll, V. Duvvuri, M. Trodden and M. Turner, {\it Physical Review D} {\bf 70}  (2004) 043528.

\bibitem{vdW}S. Capozziello, S. De Martino and M. Falanga, {\it Physics Letters A} {\bf 299} (2002) 494.

\bibitem{fR1}S. Capozziello, V. Cardone and A. Troisi, {\it Physical Review D} {\bf 71} (2005) 043503.

\bibitem{fR2}G. Allemandi, A. Borowiec and M. Francaviglia, {\it Physical Review D} {\bf 70} (2004) 103503.

\bibitem{felice} A. D. Felice and S. Tsujikawa, {\it Living Reviews in Relativity} {\bf 13} (2010) 3.

\bibitem{GB}D. Lovelock, {\it Journal Mathematical Physics} {\bf 12} (1971) 498.

\bibitem{DMfR}S. Capozziello, V. F.~Cardone and A. Troisi, {\it Monthly Notices of the Royal Astronomical Society} {\bf 375} (2007) 1423.

\bibitem{ClustersfR}S. Capozziello, E. De Filippis and V. Salzano, {\it Monthly Notices of the Royal Astronomical Society} {\bf 394} (2009) 947.

\bibitem{capoexp}S. Capozziello, V. F. Cardone, S. Carloni and A. Troisi, {\it International Journal of Modern Physics D} {\bf 12} (2003) 1969.

\bibitem{Staro}A. Starobinsky, {\it Physics Letters B} {\bf 91} (1980) 99.

\bibitem{PPN}T. Chiba, T.L. Smith and A. L. Erickcek, {\it Physical Review D} {\bf 75} (2007) 124014.

\bibitem{Lobo}O. Bertolami, C. Boehmer, T. Harko and F. Lobo, {\it Physical Review D} {\bf 75} (2007) 104016.


\bibitem{analogyf2}O. Bertolami and J. P\'aramos, {\it Classical and Quantum Gravity} {\bf 25} (2008) 245017.

\bibitem{fluid}O. Bertolami, F. S. N. Lobo and J. P\'aramos, {\it Physical Review D} {\bf 78} (2008) 064036.

\bibitem{mimic}O. Bertolami and J. P\'aramos, {\it Journal of Cosmology and Astroparticle Physics} {\bf 1003} (2010) 009. 

\bibitem{clustersBFP}O. Bertolami, P. Fraz\~ao and J. P\'aramos, {\it Physical Review D} {\bf 86} (2012) 044034.

\bibitem{accexp}O. Bertolami, P. Fraz\~ao, and J. P\'aramos, {\it Physical Review D} {\bf 81} (2010) 104046.

\bibitem{preheatingf2}O. Bertolami, P. Fraz\~ao and J. P\'aramos, {\it Physical Review D} {\bf 83} (2011) 044010.

\bibitem{perturbations}O. Bertolami, P. Fraz\~ao and J. P\'aramos, {\it Journal of Cosmology and Astroparticle Physics} {\bf 1305} (2013) 029.

\bibitem{solarBMP}O. Bertolami, R. March and J. P\'aramos, {\it Physical Review D} {\bf 88} (2013) 064019.

\bibitem{f2collapse}J. P\'aramos and C. Bastos, {\it Physical Review D} {\bf 86} (2012) 103007.

\bibitem{hydro}O. Bertolami and J. P\'aramos, {\it Physical Review D} {\bf 77} (2008) 084018.
  

\bibitem{Newtonian}O.~Bertolami and A.~Martins, {\it Physical Review D} {\bf 85} (2012) 024012.

\bibitem{BS}O. Bertolami and M. C. Sequeira, {\it Physical Review D} {\bf 79} (2009) 104010.

\bibitem{wormhole}O. Bertolami and R. Z. Ferreira, {\it Physical Review D} {\bf 85} (2012) 104050.

\bibitem{localCC}O. Bertolami and J. P\'aramos, {\it Physical Review D} {\bf 84} (2011) 064022.

\bibitem{Damour}T. Damour and G. Esposito-Far{\`e}se, {\it Classical and Quantum Gravity} {\bf 9} (1992) 2093.

\bibitem{conformal} V. Faraoni, E. Gunzig and P. Nardone, {\it  Fundamentals of Cosmic Physics} {\bf 20} (1999) 121.

\bibitem{Sotiriou1}T. P. Sotiriou and V. Faraoni, {\it Classical and Quantum Gravity} {\bf 25} (2008) 5002.

\bibitem{Seliger}R. L. Seliger and G. B. Whitham, {\it Proceedings of the Royal Society of London A} {\bf 305} (1968) 1.

\bibitem{Schutz}B. F. Schutz, {\it Physical Review D} {\bf 2} (1970) 2762.

\bibitem{Brown}J. D. Brown, {\it Classical and Quantum Gravity} {\bf 10} (1993) 1579.

\bibitem{HE}S. W. Hawking and G.F.R. Ellis, {\it The Large Scale Structure of Spacetime}, (Cambridge University Press, Cambridge 1973).

\bibitem{DK}A. D. Dolgov and M. Kawasaki, {\it Physics Letters B} {\bf 573} (2003) 1.

\bibitem{kronawitter}O. Gerhard, A. Kronawitter, R. P. Saglia and R. Bender, {\it Astrophysical Journal Supplement} {\bf 144} (2000) 53.

\bibitem{Hernquist}L. Hernquist, {\it Astrophysical Journal} {\bf 356} (1990) 359.

\bibitem{NFW}J. F. Navarro, C. S. Frenk and S. D. M. White, {\it Monthly Notices of the Royal Astronomical Society} {\bf 275} (1995) 56.

\bibitem{vikhlinin}A. Vikhlinin, A. Kravtsov, W. Forman, C. Jones, M. Markevitch, S. S. Murray and L. V. Speybroeck, {\it Astrophysical Journal} {\bf 640} (2006) 691.

\bibitem{preheating}S.~Tsujikawa, K.~i.~Maeda and T.~Torii, {\it Physical Review D} {\bf 60} (1999) 123505.

\bibitem{mimoso}T. Charters, A. Nunes and J. P. Mimoso, {\it Physical Review D} {\bf 78} (2008) 083539.

\bibitem{kinetic}J.~Lachapelle and R.~H.~Brandenberger, {\it Journal of Cosmology and Astroparticle Physics} {\bf 0904} (2009) 020.

\bibitem{WMAP7}D. N. Spergel {\it et al.}, {\it Astrophysical Journal Supplement} {\bf 170} (2007) 377. 

\bibitem{gong}Y. G. Gong and A. Wang, {\it Physical Review D} {\bf 75} (2007) 043520.

\bibitem{pert11}T. Koivisto, {\it Physical Review D} {\bf 73} (2006) 083517.

\bibitem{pert12}R. Bean {\it et al}., {\it Physical Review D} {\bf 75} (2007) 064020.

\bibitem{pert13}B. Li and J. D. Barrow, {\it Physical Review D} {\bf 75} (2007) 084010.

\bibitem{pert14}S. Carloni, P. K. S. Dunsby and A. Troisi, {\it Physical Review D} {\bf 77} (2008) 024024.

\bibitem{pert15}K. Bamba {\it et al.}, {\it Astrophysics and Space Science}  {\bf 342} (2012) 155.

\bibitem{Zhang}P. Zhang {\it et al.}, {\it Physical Review Letters} {\bf 99} (2007) 141302.

\bibitem{Mukhanov}V. F.~Mukhanov, H. A. Feldman and R. H. Brandenberger, {\it Physical Reports} {\bf 215} (1992) 203.

\bibitem{Cruz}A. de la Cruz-Dombriz, A. Dobado and A. L. Maroto, {\it Physical Review D} {\bf 77} (2008) 123515.

\bibitem{weaklensing1}V. Acquaviva, C. Baccigalupi and F. Perrotta, {\it Physical Review D} {\bf 70} (2004) 023515.

\bibitem{weaklensing2}S. Tsujikawa and T. Tatekawa, {\it Physics Letters B} {\bf 665} (2008) 325.

\bibitem{Will}C. Will, {\it Living Reviews in Relativity} {\bf 9} (2005) 3.

\bibitem{beta}E.~G. Adelberger, {\it Classical and Quantum Gravity} {\bf 18} (2001) 2397.

\bibitem{gamma}B. Bertotti, L. Iess and P. Tortora, {\it Nature} {\bf 425} (2003) 374.

\bibitem{survive}O. Bertolami, J. P\'aramos and S. G. Turyshev, ``General theory of relativity: Will it survive the next decade?" in Lasers, Clocks, and Drag-Free: Technologies for Future Exploration in Space and Tests of Gravity; H. Dittus, C. Laemmerzahl, S. Turyshev, Eds. (Springer Verlag, 2006); arXiv:0602016 [gr-qc].

\bibitem{status} O. Bertolami and J. P\'aramos, ``The experimental status of Special and General Relativity'', in Handbook of Spacetime; A. Ashtekar, V. Petkov. Eds. (Springer Verlag, 2014); arXiv:1212.2177 [gr-qc].

\bibitem{PPNfR}S. Capozziello, M. De Laurentis, S. Nojiri and S. D. Odintsov, {\it General Relativity and Gravitation} {\bf 41} (2009) 2313.

\bibitem{NASAdensity}NASA website, {\it http://spacemath.gsfc.nasa.gov}.

\bibitem{Sotiriou3}T. P. Sotiriou, {\it Physics Letters B} {\bf 664} (2008) 225.

\bibitem{bullet}D. Clowe, M. Bradac, A. Gonzalez, M. Markevitch, S. Randall, C. Jones and D. Zaritsky, {\it Astrophysical Journal} {\bf 648} (2006) L109.

\end{thebibliography}
\end{document}